\documentclass{appolb}

\usepackage{latexsym}
\usepackage{amsmath}
\usepackage{amssymb}
\usepackage{amsfonts}
\usepackage{bm}
\usepackage{physics}
\usepackage[table,x11names]{xcolor}
\usepackage{color}
\definecolor{purple}{rgb}{0.5,0,0.5}
\definecolor{blue}{rgb}{0.0,0,0.9}
\definecolor{prdblue}{rgb}{0.133,0.118,0.498}
\usepackage[colorlinks=true, pdfstartview=FitV, linkcolor=prdblue, citecolor= prdblue, urlcolor=prdblue]{hyperref}

\usepackage{supertabular} 
\usepackage{placeins}
\usepackage{epsfig}

\usepackage{graphicx}

\begin{document}
\title{A Continuum Schwinger Method to Study the Pion's Generalized Parton Distribution%
\thanks{Presented at the Excited QCD 2026 Workshop by J. Segovia}%
}

\author{J.M. Morgado-Ch\'avez
\address{Universidad de Valencia and IFIC-CSIC, E-46100, Valencia, Spain}
\\[3mm]
J. Segovia, F. de Soto
\address{Universidad Pablo de Olavide, E-41013 Sevilla, Spain}
\\[3mm]
J. Rodríguez-Quintero
\address{Universidad de Huelva, E-21071 Huelva, Spain}
\\[3mm]
V. Bertone, M. Defurne, C. Mezrag, H. Moutarde
\address{Irfu, CEA, Universit\'e Paris-Saclay, 91191, Gif-sur-Yvette, France}
}

\maketitle

\begin{abstract}
Generalised Parton Distributions (GPDs) provide multidimensional insight into hadron structure and are particularly relevant for the pion, whose dynamics are intimately linked to chiral symmetry breaking. We introduce a novel modelling strategy for pion GPDs that satisfies all QCD constraints by construction: support, polynomiality, positivity, and the soft-pion theorem. The approach is illustrated with a simple algebraic model, which is evolved and used to compute deeply virtual Compton scattering (DVCS) Compton Form Factors at next-to-leading order. Our results indicate that gluons dominate the pion response at the Electron Ion Collider kinematics.
\end{abstract}

\vspace*{-0.40cm}

\section{Introduction}
\label{sec:Introduction}

Describing hadron structure in terms of quark and gluon degrees of freedom remains a central challenge in strong-interaction physics. Upcoming experimental facilities dedicated to this goal~\cite{AbdulKhalek:2021gbh} have stimulated extensive theoretical developments~\cite{Diehl:2003ny}.

Among hadrons, the pion plays a special role as the Nambu--Goldstone boson associated with dynamical chiral symmetry breaking in QCD~\cite{Aguilar:2019teb}. Understanding how quarks and gluons generate its structure is therefore essential for clarifying the emergence of mass in the Standard Model.

We study whether future electron–ion colliders can access pion structure via GPDs using the Sullivan process. After outlining the theoretical framework, we introduce a modelling strategy that fulfills QCD constraints and compute DVCS Compton Form Factors to assess the gluon contribution at collider kinematics.

\vspace*{-0.40cm}

\section{Sullivan process: one-pion-exchange approximation}
\label{sec:DVCS}

Deep inelastic lepton-hadron scattering in the Bjorken limit provides access to partonic degrees of freedom. While inclusive DIS constrains parton distribution functions, some exclusive channels, \emph{e.g.} DVCS, allow one to probe Generalised Parton Distributions (GPDs).

Following Sullivan~\cite{Sullivan:1971kd}, we consider the one-pion-exchange contribution to deep inelastic electron-proton scattering with a tagged neutron in the final state. For small momentum transfer between the initial proton and final neutron, the amplitude is dominated by pion exchange. In this kinematic regime, the process can be interpreted as scattering off a nearly on-shell virtual pion.

To suppress resonance contributions, one requires the invariant mass of the hadronic system to satisfy $W^2 \gtrsim 4\,\text{GeV}^2$~\cite{Amrath:2008vx}. Besides, one works at low momentum transfer, $|t|\lesssim 0.6\,\text{GeV}^2$~\cite{Qin:2017lcd}, near the pion pole to ensure dominance of the one-pion-exchange mechanism. And, finally, selecting longitudinal photon polarization further ensures the factorization scheme~\cite{Collins:1998be}.

Under these conditions, two mechanisms contribute: (i) deeply virtual Compton scattering (DVCS), sensitive to pion GPDs and thus to its three-dimensional structure; (ii) the Bethe--Heitler process, which depends on the pion electromagnetic form factor.

Since high-energy pion electro-production has already been used to extract the pion form factor at large $Q^2$~\cite{JeffersonLab:2008jve}, it is natural to ask whether the DVCS channel in the Sullivan process can provide access to pion GPDs.

\vspace*{-0.40cm}

\section{Generalised Parton Distributions}
\label{sec:GPDs}

GPDs are defined as matrix elements of non-local light-cone operators between hadronic states with different momenta and helicities~\cite{Ji:1996nm}. They depend on three kinematic variables:
\begin{itemize}
\item $x$, the average light-cone momentum fraction of the active parton;
\item $\xi$, the longitudinal momentum transfer (skewness);
\item $t$, the invariant momentum transfer.
\end{itemize}

GPDs interpolate between parton distribution functions and form factors and evolve with the renormalization scale $\mu$. Two kinematic regions are distinguished: the DGLAP region $(|x|\ge|\xi|)$ and the ERBL region $(|x| \le |\xi|)$.

In impact-parameter space, GPDs admit a probabilistic interpretation as spatial distributions of partons in the transverse plane~\cite{Burkardt:2000za}, providing access to the multidimensional structure of hadrons.

Any realistic pion GPD model must satisfy fundamental constraints imposed by QCD~\cite{Diehl:2003ny}:
\begin{itemize}
\item Support: $(x,\xi) \in [-1,1] \times [-1,1]$, as required by causality and analyticity.

\item Polynomiality: the $m$-th Mellin moment is a polynomial in $\xi$ of degree $m+1$, reflecting Lorentz invariance.

\item Positivity: in the DGLAP region, GPDs are bounded by PDFs through the Cauchy--Schwarz inequality,
\begin{equation}
\label{eq:PositivityBound}
\left|H^{q}_{\pi^{+}}\left(x,\xi,t;\mu\right)\right|\leq\sqrt{q_{\pi^{+}}\left(\frac{x+\xi}{1+\xi};\mu\right)q_{\pi^{+}}\left(\frac{x-\xi}{1-\xi};\mu\right)} \,\,\,\, \left|x\right|\geq\left|\xi\right| \,,
\end{equation}

\item Soft-pion theorem: axial Ward--Takahashi identities and PCAC constrain the isoscalar and isovector combinations of pion GPDs~\cite{Polyakov:2002yz}.
\end{itemize}

These requirements severely restrict model building and must be implemented consistently for reliable phenomenology.

\vspace*{-0.40cm}

\section{GPD modelling}
\label{sec:Modeling}

Constructing pion GPDs that simultaneously satisfy support, polynomiality, positivity, and the soft-pion theorem remains a nontrivial task. We introduce a modelling strategy that enforces these constraints by construction and relies on a single dynamical input: the pion light-front wave function (LFWF).

The procedure consists of three steps:
\begin{enumerate}
\item Light-front input. Start from a factorised pion LFWF at a reference scale $\mu_{\text{ref}}$,
\begin{equation}
\Psi^{q}_{\pi^+}(x,k_\perp;\mu_{\text{ref}}) = \mathcal{N}_\Psi \, \phi(x;\mu_{\text{ref}}) \, \varphi(k_\perp;\mu_{\text{ref}}) \,,
\end{equation}
separating longitudinal and transverse dynamics.

\item DGLAP GPD from overlap. Using the overlap representation~\cite{Hwang:2007tb}, construct the GPD in the DGLAP region:
\begin{align}\label{eq:GPDmodel}
H^{q}_{\pi^{+}}\left(x,\xi,t;\mu_{\text{ref}}\right)\Big|_{\left|x\right|\geq\left|\xi\right|} &= \mathcal{N}_{H} \sqrt{q_{\pi^{+}}\left(\frac{x+\xi}{1+\xi};\mu_{\text{ref}}\right) q_{\pi^{+}}\left(\frac{x-\xi}{1-\xi};\mu_{\text{ref}}\right)} \nonumber\\
&
\times \Phi\left(x,\xi,t;\mu_{\text{ref}}\right) \,.
\end{align}
Canonical normalisation implies $\Phi(x,\xi,0)=1$, so positivity is saturated in the DGLAP region.

\item ERBL completion. The ERBL region is obtained via the covariant extension~\cite{Chouika:2017dhe}, which guarantees polynomiality. The soft-pion theorem then fixes the chiral limit, ensuring consistency with axial Ward--Takahashi identities.
\end{enumerate}

This yields pion GPDs that fulfill all QCD constraints and are fully determined once the pion parton distribution function is specified~\cite{Chavez:2021llq, Chavez:2021koz}.

The assumption of $x$-$k_\perp$ factorization can be justified using perturbation-theory integral representations of pion Bethe--Salpeter amplitudes~\cite{Mezrag:2016hnp}. In the isospin-symmetric limit, the transverse component acquires a simple dipole form, leading to an intrinsic and controlled $t$-dependence. Consequently, the only remaining modelling freedom resides in the choice of the pion PDF. As an illustration, consider the simple ansatz for the pion PDF~\cite{Chouika:2017dhe}:
\begin{equation}\label{eq:PDF}
q_{\pi^{+}}\left(x;\mu_{\text{Ref.}}\right)=30x^{2}\left(1-x\right)^{2} \,.
\end{equation}
Note that more sophisticated PDFs derived using continuum Schwinger methods have been employed in Refs.~\cite{Chavez:2021llq, Chavez:2021koz}.

\vspace*{-0.40cm}

\section{DVCS Compton Form Factors (CFFs)}
\label{sec:CFFs}

Deeply virtual Compton scattering (DVCS) amplitudes are parametrised in terms of Compton Form Factors (CFFs), obtained as convolutions of GPDs with perturbatively calculable kernels~\cite{Ji:1996nm}:
\begin{equation}
\mathcal{H}_{\pi^{+}}\left(\xi,t;Q^{2}\right)=\sum_{p=\left\lbrace q\right\rbrace,g}\int_{-1}^{1}\frac{dx}{\xi}\mathcal{K}^{p}\left(\frac{x}{\xi},\frac{Q^{2}}{\mu^{2}_{\text{F}}},\alpha_{\text{S}}\left(\mu_{\text{F}}^{2}\right)\right)H^{p}_{\pi^{+}}\left(x,\xi,t;\mu_{F}^{2}\right) \,.
\end{equation}

To confront experimental kinematics, the GPDs must be evolved from the reference scale $\mu_{\text{ref}}=0.331$~[27] to the relevant factorisation scale using QCD evolution equations. This is implemented with Apfel$++$~\cite{Bertone:2017gds}, while NLO CFFs are computed within the PARTONS framework~\cite{Berthou:2015oaw}.

\begin{figure}[!t]
\centering
\includegraphics[scale=0.4]{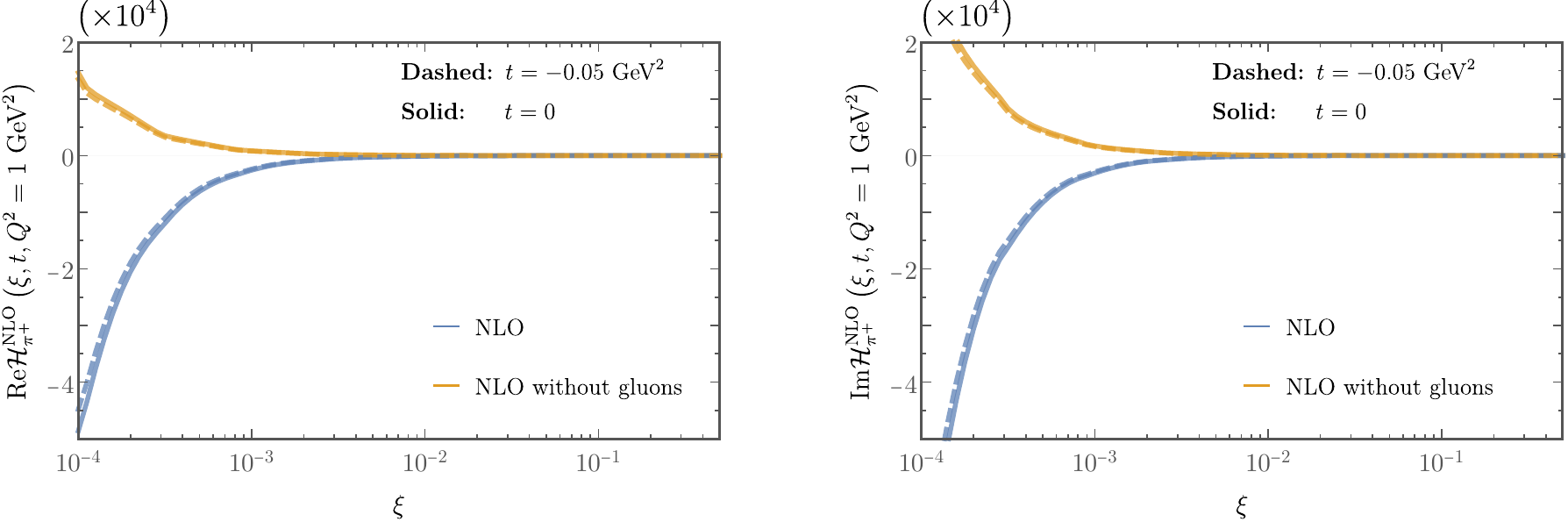}
\caption{NLO DVCS Compton form factors yield by the algebraic model of Eq.~\eqref{eq:PDF}. Yellow plots correspond to the NLO calculation with the input gluon distribution set to zero.}
\label{fig:CFF}
\end{figure}

The resulting NLO CFFs (Fig.~\ref{fig:CFF}) show a clear dominance of gluon contributions at collider scales. Setting the gluon distribution to zero leads to a substantial suppression of both real and imaginary parts of the CFFs. Hence, gluons control the pion response in DVCS within the one-pion-exchange regime, indicating that future EIC measurements will primarily probe gluonic structure.

\vspace*{-0.60cm}

\section{Conclusions}

The Sullivan process provides a promising avenue to access pion Generalised Parton Distributions at future electron-ion colliders. We have presented a modelling strategy that enforces all QCD constraints by construction, relying solely on the pion light-front wave function, and its associated PDF. Our analysis shows that gluons dominate the pion contribution to DVCS Compton Form Factors at collider scales, highlighting the central role of gluonic dynamics in high-energy pion structure and the strong sensitivity of future measurements to this sector.

\vspace*{0.40cm}
{\bf Acknowledgements.} Work supported by MINCIN under grant No. PID2019-107844GB-C22; Junta de Andalucía under contract No. PAIDI FQM-370; EU Horizon 2020 research and innovation programme under grant agreement No. 824093; and through the GLUODYNAMICS project funded by ``P2IO LabEx (ANR-10-LABX-0038)" in the framework ``Investissements d'Avenir" managed by the ANR, France.


\bibliographystyle{ieeetr} 
\bibliography{printCSM-PionGPD-eQCD}

\end{document}